\documentclass[letterpaper]{jpconf}
\usepackage{graphicx}
\usepackage{color}

             \font\sevenrm=cmr7

          \font\sixrm=cmr6

\def\erg{\varepsilon}

\def\rns{R_{\hbox{\sixrm NS}}}

\def\rmax{r_{\rm max}}

\def\sigt{\sigma_{\hbox{\sixrm T}}}

\def\ThetaBn{\Theta_{\hbox{\sevenrm B}n}}

\def\emph#1{{\it #1}}

{\catcode`\@=11                                                  
\gdef\SchlangeUnter#1#2{\lower2pt\vbox{\baselineskip 0pt\lineskip0pt    
\ialign{$\m@th#1\hfil##\hfil$\crcr#2\crcr\sim\crcr}}}}           
\def\gtrsim{\mathrel{\mathpalette\SchlangeUnter>}}               
\def\lesssim{\mathrel{\mathpalette\SchlangeUnter<}}    
\def\teq#1{$\, #1\,$}                         

\begin{document}

\newcommand{\vol}[2]{$\,$\bf #1\rm , #2.}                 

\noindent
{\small To appear in Proc. ``Physics of Neutron Stars -- 2017,'' Journal of Physics: Conference Series,
eds. G.~G. Pavlov, J.~A. Pons, P.~S. Shternin, D.~G. Yakovlev,
held in Saint Petersburg, Russia, 10--14 July, 2017}
\vspace{-20pt}

\title{Hard X-ray Quiescent Emission in Magnetars via Resonant Compton Upscattering}

\author{M.~G.\ Baring$^1$, Z. Wadiasingh$^2$, P.~L. Gonthier$^3$, A.~K. Harding$^4$}

\address{$^1$ Department of Physics and Astronomy,
      Rice University, Houston, TX 77251, U.S.A.}
\address{$^2$ Centre for Space Research,
   	North-West University, Potchefstroom, South Africa}
\address{$^3$ Hope College, Department of Physics,
          27 Graves Place, Holland, MI 49423, U.S.A}
\address{$^4$ Astrophysics Science Division, Code 663
	NASA's Goddard Space Flight Center, Greenbelt, MD 20771, U.S.A.}

\ead{baring@rice.edu}

\begin{abstract}
Non-thermal quiescent X-ray emission extending between 10 keV and around 150 keV
has been seen in about 10 magnetars by RXTE, INTEGRAL, Suzaku, NuSTAR and
Fermi-GBM. For inner magnetospheric models of such hard X-ray signals, inverse
Compton scattering is anticipated to be the most efficient process for
generating the continuum radiation, because the scattering cross section is
resonant at the cyclotron frequency. We present hard X-ray upscattering spectra
for uncooled monoenergetic relativistic electrons injected in inner regions of
pulsar magnetospheres.  These model spectra are integrated over bundles of
closed field lines and obtained for different observing perspectives.  The
spectral turnover energies are critically dependent on the observer viewing
angles and electron Lorentz factor.  We find that electrons with energies less
than around 15 MeV will emit most of their radiation below 250 keV, consistent
with the turnovers inferred in magnetar hard X-ray tails.  Electrons of higher
energy still emit most of the radiation below around 1 MeV, except for
quasi-equatorial emission locales for select pulse phases. Our spectral
computations use a new state-of-the-art, spin-dependent formalism for the QED
Compton scattering cross section in strong magnetic fields.
\end{abstract}

\section{Introduction}
A topical subset of the neutron star population is defined by magnetars. They
are highly-magnetized stars that have historically been divided into two
observational groups: Soft-Gamma Repeaters (SGRs) and Anomalous X-ray Pulsars
(AXPs).  Their extreme fields, which generally exceed the quantum critical value
of \teq{B_{\rm cr}=m_e^2c^3/(e\hbar)\approx 4.41\times 10^{13}}Gauss (where the
electron cyclotron and rest mass energies are equal), are inferred from their
X-ray timing properties, presuming that their rapid rotational spin down is due
to magnetic dipole torques (e.g. \cite{kouv98}).  Such a class of neutron stars
with superstrong fields was postulated for SGRs by \cite{dt92}, and for AXPs by
\cite{td96}.  For recent reviews of magnetar science, see
\cite{tzw15,kb17,CotiZelati17}.

SGRs are transient sources that undergo repeated hard X-ray outbursts. Most of
their ephemeral activity consists of short flares of subsecond duration in the
\teq{10^{38}}erg/sec \teq{ < L < 10^{42}} erg/sec range, often somewhat isolated
in time, and sometimes occurring in storms. Yet three magnetars have exhibited
giant supersecond flares of energies exceeding \teq{10^{45}}ergs: SGR 0525-66 in
1979, as mentioned above, SGR 1900+14 on August 27th, 1998 (e.g. see
\cite{hurl99}), and SGR 1806-20 on December 27, 2004 (see \cite{palm05}). SGRs
also exhibit quiescent X-ray emission below \teq{10}keV, of periods
\teq{P} in the range 2--12 sec. (e.g. \cite{kouv98,kulk03}). The AXPs are bright
X-ray sources with periods in the same range.  Their quiescent signals below 10
keV are mostly thermal with steep power-law tails, and possess luminosities
\teq{L_X \sim 10^{33}-4 \times 10^{35}\,\rm erg\; s^{-1}} \cite{vigano13}.\footnote{See
also the neutron star cooling site {\tt http://neutronstarcooling.info/} for
magnetars in a broader context.} As with the SGRs, these \teq{L_X} values far
exceed their rotational power, perhaps being powered by their internal magnetic
energy.  Observations of outburst activity in AXP 1E 2259+586 \cite{kaspi03}, in
AXP 1E1841-045 \cite{lin11} and in others suggest that AXPs and SGRs are indeed
very similar. This ``unification paradigm'' has garnered widespread support
within the magnetar community over the last decade. There are also
highly-magnetized pulsars that exhibit periods of magnetar-like bursting
activity. The observational status quo of magnetars is summarized in the McGill
Magnetar Catalog \cite{ok14}.\footnote{An on-line version can be found at {\tt
http://www.physics.mcgill.ca/\~{}pulsar/magnetar/main.html}}

An additional element of the magnetar phenomenon emerged following the discovery
by INTEGRAL and RXTE of hard, non-thermal pulsed spectral tails in three AXPs
\cite{kuip04,hartog08}. These luminous tails are extremely flat, extending up to
150 - 200 keV. Similar quiescent emission tails are seen in SGRs
\cite{goetz06,enoto10}.  In various magnetars, a turnover around 500 - 750 keV
is implied by constraining {\it pre-2000} upper limits obtained by the COMPTEL
instrument on the \emph{Compton Gamma-Ray Observatory}. The need for such a
spectral turnover is reinforced above 100 MeV by upper limits in
\emph{Fermi}-LAT data for around 20 magnetars \cite{abdo10,li17}. Magnetic
inverse Compton scattering of thermal atmospheric soft X-ray seed photons by
relativistic electrons is expected to be extremely efficient in magnetars, and
thus is a prime candidate for generating the hard X-ray tails \cite{bh07,ft07}.
This paper explores this model, highlighting some recent results from our
ongoing program of modeling the resonant Compton hard X-ray emission in
magnetars.

\section{Hard X-rays in Magnetars from Magnetic Inverse Compton Scattering}
The scenario for the generation hard X-ray tails considered in this paper is
magnetic inverse Compton scattering of thermal atmospheric soft X-ray seed
photons by relativistic electrons.  This is extremely efficient in
highly-magnetized pulsars because the scattering process is resonant at the
electron cyclotron frequency and its harmonics, so that there the cross section
in the electron rest frame exceeds the classical Thomson value of \teq{\sigt
\approx 6.65 \times 10^{-25}}cm$^2$ by \teq{\sim 2-3} orders of magnitude (e.g.,
\cite{dh86,gonthier00}). This efficiency is manifested in short cooling lengths,
often less than \teq{10^3}cm, for high speed electrons traversing a magnetar
magnetosphere \cite{bwg11}. Provided there is a source of electrons with Lorentz
factors $\gamma_e \gg 1$, single inverse Compton scattering events can readily
produce the general character of hard X-ray tails \cite{bh07,ft07}. In
particular, Baring \& Harding \cite{bh07} employed QED scattering cross sections
in uniform fields, extending collision integral formalism for non-magnetic
Compton upscattering that was developed by \cite{he89}.

The spectra presented in \cite{bh07} were characteristically flat, a consequence
of the resonant cyclotron kinematics.  These do not match observations, nor are
they expected to since they integrate over all lines of sight in the uniform
{\bf B}.  Non-uniform fields offer a different weighting of angular geometries,
and when combined with cooling can steepen the spectrum considerably: see the
magnetic Thomson investigation of \cite{Beloborodov13}.   In \cite{bh07}, we
discerned that kinematic constraints correlating the directions and energies of
upscattered photons yielded Doppler boosting and blueshifting along the local
magnetic field direction.  Therefore, the strong angular dependence of spectra
computed for the uniform field case must extend to more complex magnetospheric
field configurations. Consequently, emergent inverse Compton spectra in more
complete models of hard X-ray tails will depend critically on an observer's
perspective and the sampled locales of resonant scattering, both of which vary
with the rotational phase of a magnetar.   The construction of the resonant
Compton upscattering model whose results are presented here is a geometrical
extension of the work of \cite{bh07} to dipolar magnetic field morphologies.
Directed emission spectra have been generated for an array of observer
perspectives and magnetic inclination angles \teq{\alpha} to the rotation axis;
they serve as a basis for future calculations that will treat Compton cooling of
electrons self-consistently. The scattering physics employs the
state-of-the-art, spin-dependent magnetic Compton formalism developed by us in
Gonthier et al. \cite{gonthier14} that uses the preferred Sokolov and Ternov
eigenstates of the Dirac equation. For details of the model, its kinematics and
geometry and spectral characteristics, and their connection to observer
perspectives, the reader is referred to the full exposition in Wadiasingh et al.
\cite{wbgh17}.

\begin{figure}[h]
\begin{center}
\includegraphics[width=\textwidth]{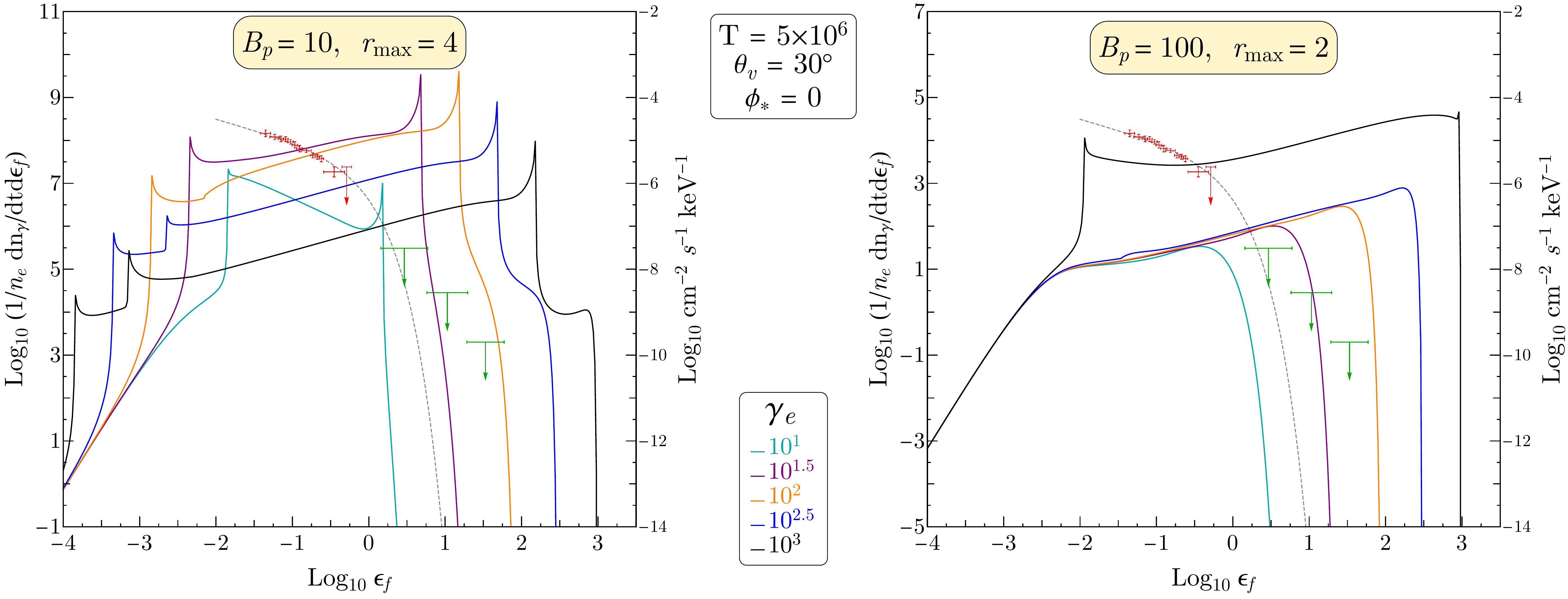}
\end{center}
\vspace{-18pt}
\caption{\label{fig:spectra}
Spectra generated for  meridional field loops, those where the line of sight to 
an observer is coplanar ($\phi_* =0$) with the loops.  The viewing angle of the 
observer is $\theta_v = 30^\circ$ from the magnetic dipole axis.  Results for various
electron Lorentz factors are depicted.  The left panel illustrates higher-altitude 
and lower-field directed spectra computed for $B_{\rm p} = 10$ and $\rmax = 4$ 
where the resonant interactions are readily sampled for Lorentz factors 
$\gamma_e>10^2$. For the right panel with $\rmax =2$ and  $B_{\rm p} = 100$, 
the local field is much higher, precluding resonant interactions near equatorial 
regions unless Lorentz factors are much higher.  Overlaid on the 
computed spectra are observational data points for AXP 4U 0142+61 
(den Hartog et al. 2008b) along with a schematic \teq{\erg_f^{-1/2}} power-law 
with a \teq{250} keV exponential cutoff (black dotted curve).}
\end{figure}

Two elements of the extensive work in \cite{wbgh17} are highlighted here.  The
first consists of selected but informative spectra, depicted in
Fig.~\ref{fig:spectra}, computed for electrons of fixed Lorentz factors
traversing individual field lines.  They correspond to viewing angles coplanar
with the field loops (meridional cases) that readily sample the Doppler boosting
and beaming. The combination of \teq{\gamma_e} and local field \teq{B \sim
B_{\rm p}/\rmax^3 \sim \gamma_e \erg_f (1 + \cos \ThetaBn)} essentially controls
access to resonant interactions \cite{bh07}, and the value of the spectral
index.  Here \teq{\erg_f} is the upscattered photon energy in units of
\teq{m_ec^2}, and \teq{\ThetaBn} is the observer's angle relative to {\bf B} at
the point of scattering. The coupling \teq{B \sim \gamma_e \erg_f (1 + \cos
\ThetaBn)} controls the directionality of emitted photons, with higher energies
\teq{\erg_f} being beamed closer to the field lines.  For much of the range in
\teq{\erg_f}, the spectra that sample resonant interactions possess a
characteristic scaling \teq{dn/(dt d\erg_f) \sim \erg_f^{1/2}}, i.e. are
extremely flat, even harder than the uniform field results in \cite{bh07}.  This
approximate power-law dependence is a consequence of kinematics and
magnetospheric geometry \cite{wbgh17}.  Bracketing these quasi-power-law bands
are distinctive ``horns" or cusps, distinguishing when resonant scatterings are
and are not accessed. These appear both at low values \teq{\erg_f} in the soft
X-rays/EUV, where they would be dominated by the surface emission signal (not
shown), and also in the hard X-ray and gamma-ray domains.  The narrow peaks of
the horns are weighted images of the resonant differential cross section,
enhanced by the beaming. Not all spectra possess frequency ranges where resonant
interactions are accessible: for values of local \teq{B} that are large,
resonant interactions in the Wien peak are often not fully sampled, as is
evident from computed spectra presented in the right panel of
Fig.~\ref{fig:spectra}, and the \teq{\gamma_e=10} example in the left panel.

\newpage

Also illustrated in Fig.~\ref{fig:spectra} are renormalized hard X-ray spectral
data for one magnetar, to illustrate how the monoenergetic electron model from
single field loops does not match observations.  Modeling of hot thermal surface
emission seeded by electron bombardment at loop footpoints is not undertaken. 
The computed spectra extend beyond the COMPTEL upper limits (green points) when
\teq{\gamma_e \gtrsim 30}. Therefore, lower Lorentz factors \teq{\gamma_e\sim
10} are desirable, and these a naturally generated by electron cooling
\cite{bwg11}.  High-energy attenuation mechanisms like photon splitting or
magnetic pair creation may also be operating. In \cite{wbgh17} we also exhibit
spectra from integrations over field line azimuths, encompassing the
non-meridional loops that dominate the contribution from a toroidal surface
comprising dipolar field lines.  These demonstrate steeper \teq{dn/(dt d\erg_f)
\sim \erg_f^{0}} power-laws because the loops that do not provide tangents (i.e.
{\bf B} directions) at scattering locales that point toward an observer soften
the spectrum.  Even more interestingly, \cite{wbgh17} illustrate that spectral
arrays over such toroidal surfaces that span a range of altitudes yield an
envelope that approximately matches the 4U0142+61 spectrum displayed in
Fig.~\ref{fig:spectra}, provided that \teq{\gamma_e \sim 10}. This suggests that
models with more complete volumetric integrations and electron cooling
incorporated will match the spectroscopy of hard X-ray tails.

\vspace{-7pt}
\begin{figure}[h]
\begin{center}
\includegraphics[width=\textwidth]{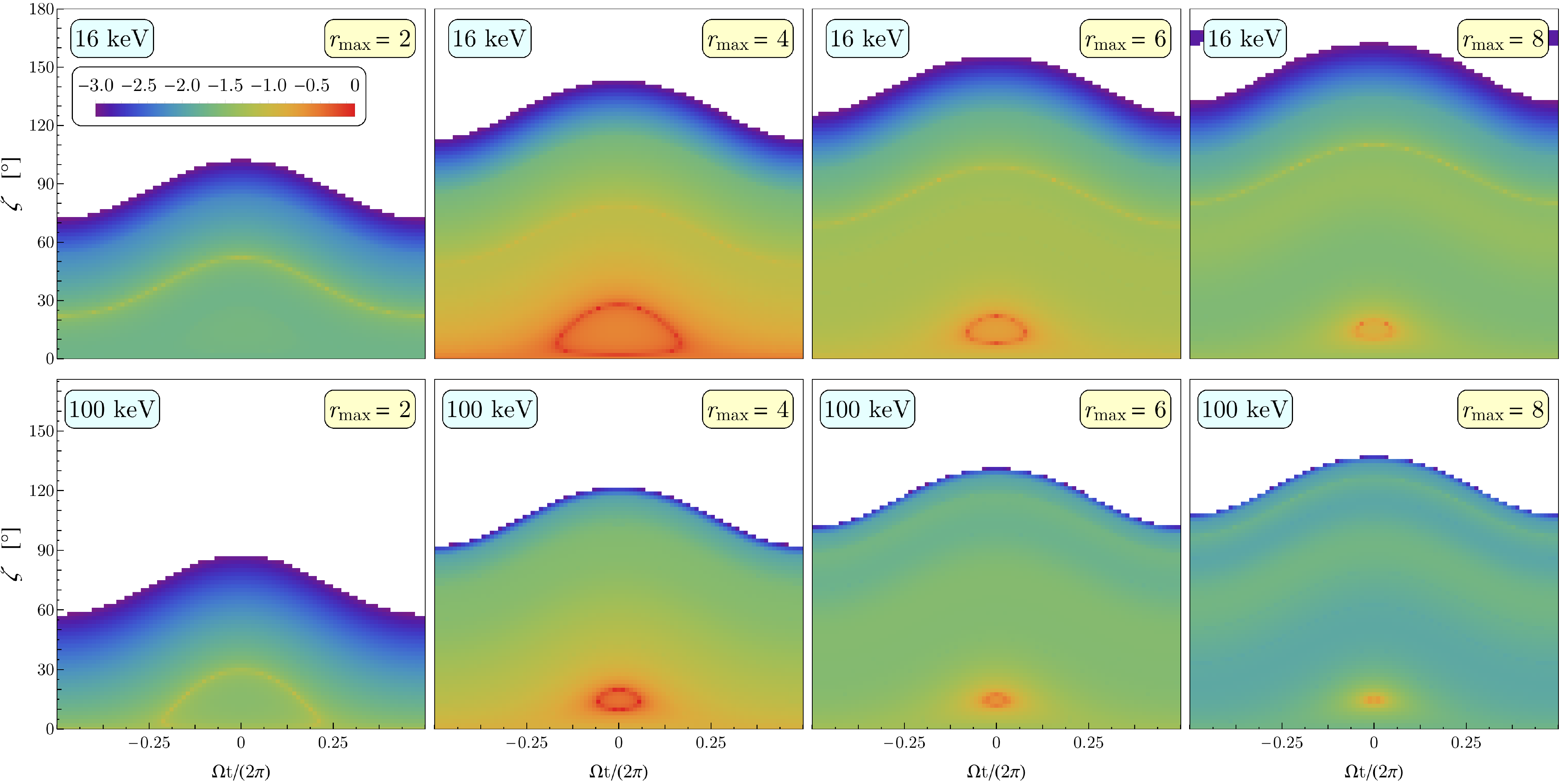}
\end{center}
\vspace{-20pt}
\caption{\label{fig:phase_plots}
Normalized photon flux \teq{\zeta - \Omega t/2\pi} phase space maps for 
resonant Compton upscattering.  These represent the 
logarithmically-scaled (base 10) intensity at energies \teq{16} keV (top row)
and \teq{100} keV (bottom row), color coded according to the legend, as a function of 
spin phase \teq{\Omega t/2\pi} for each value of \teq{\zeta} on the ordinate.  
The intensity maps are for uncooled electrons with \teq{\gamma_e = 10^2}, 
and a uniform surface temperature \teq{T=5 \times 10^6}K. 
They are obtained for azimuthally-integrated bundles 
of field lines, i.e. a toroidal surface, with \teq{B_p=10}, and stellar magnetic inclination
\teq{\alpha = 15^{\circ}}, and
depict maps for maximum loop altitudes \teq{r_{\rm max}=2,4,6,8}.
Pulse profiles for a particular observer \teq{\zeta} are represented by horizontal cuts 
of the maps.}
\end{figure}
\vspace{-7pt}

The spectral view so far has been for an instantaneous viewing perspective, i.e.
fixed pulse phases.  Another element delivered in \cite{wbgh17} was the
representation of how pulse profiles generated for spectra from toroidal
surfaces vary with photon energy \teq{\erg_f} and the maximum surface altitude
\teq{\rmax} (in units of \teq{\rns}).  These were expressed as sky maps in the 
\teq{\zeta - \Omega t/2\pi} plane, and an example is depicted in
Fig.~\ref{fig:phase_plots}.  The intensity scale is logarithmic, and is
normalized so that the maximum in each row of panels is set to unity. Here,
\teq{\zeta} is the angle between the viewer's direction and the magnetar spin
axis, and for a particular choice of this angle, horizontal sections within each
panel define an intensity trace with pulse phase. Generally, the pulsation
profiles are of smooth, single-peaked character. Yet a symmetric double-peak
structure of the profiles in domains \teq{\zeta \approx \alpha} is evident,
being manifested as sections of the red rings: these are realized when
quasi-polar viewing is possible at select phases.  The phase separation of this
double-peak structure in domains \teq{\alpha \approx \zeta} shrinks at higher
\teq{r_{\rm max}} and larger \teq{\erg_f}. This identifies a potentially potent
observational diagnostic:  comparing theoretical energy-dependent pulse profiles
with observational ones can infer values for \teq{\alpha} and \teq{\zeta} in
magnetars, an analogous protocol to that widely used in gamma-ray pulsar
studies.  To this end, \cite{wbgh17} applied this to the observed phase
separation of 0.4 between the two peaks in the pulse profile of 1E 1841-045 
within the energy range of 20-35 keV.  We found that for \teq{\gamma_e\sim 10}
that would result from strong cooling, this suggests that \teq{\alpha \lesssim
20^{\circ}}, an estimate that is quite similar to the value of \teq{\alpha \sim
15^{\circ}} inferred in the analysis of \cite{an15}.   This determination would
change if toroidal components to the equatorial field (scattering locales that
dominate the spectral signal) yield twist angles \teq{\Delta \phi \sim 1}, i.e.,
significantly larger than those found at colatitudes \teq{\theta > 60^\circ} in
MHD simulations of field untwisting \cite{cb17}.

In summary, the sample results presented here provide an idea of the constraints
imposed upon our model, and a taste of the promise of the resonant Compton
upscattering picture in explaining the phenomenon of the hard X-ray tails in
magnetar quiescent emission.  They also suggest the usefulness of the model in
probing the magnetic angle \teq{\alpha}, thus aiding in reducing uncertainties
in the determination of magnetar field strengths using spin-down information.

\vspace{-12pt}
\ack 
M.G.B. acknowledges support by the 
NASA Astrophysics Theory (ATP) and {\it Fermi} Guest Investigator Programs
through grants NNX13AQ82 and NNX13AP08G. 
Z.W. is supported by the South African National
Research Foundation. 
P.L.G. thanks the Michigan Space Grant Consortium, the National Science Foundation (grant AST-1009731), 
and NASA ATP through grant NNX13AO12G.
A.K.H. also acknowledges support through the NASA ATP
program.

\vspace{-10pt}
\section*{References}

\def\mnras{M.N.R.A.S.}
\def\aassupp{{Astron. Astrophys. Supp.}}
\def\apss{{Astr. Space Sci.}}
\def\apj{ApJ}
\def\nat{Nature}
\def\aaps{{Astron. \& Astr. Supp.}}
\def\aap{{A\&A}}
\def\apjs{{ApJS}}
\def\sp{{Solar Phys.}}
\def\jgr{{J. Geophys. Res.}}
\def\jphysb{{J. Phys. B}}
\def\ssr{{Space Science Rev.}}
\def\araa{{Ann. Rev. Astron. Astrophys.}}
\def\nature{{Nature}}
\def\asr{{Adv. Space. Res.}}
\def\rmp{{Rev. Mod. Phys.}}
\def\prc{{Phys. Rev. C}}
\def\prd{{Phys. Rev. D}}
\def\pr{{Phys. Rev.}}

\end{document}